\documentstyle[twocolumn,psfig,aps]{revtex}
\begin{document}
\def\half{\frac{1}{2}}

\draft
\title{Diffusons, Locons, Propagons: Character of Atomic Vibrations 
in Amorphous Si}

\author{ Philip B. Allen}
\address{Department of Physics and Astronomy, State University of New York,
Stony Brook, New York 11794-3800}
 
\author{Joseph L. Feldman}
\address{Naval Research Laboratory, Washington DC 20375-5345}

\author{Jaroslav Fabian}
\address{Department of Physics, University of Maryland at College Park,
College Park, Maryland 20742-4111}

\author{Frederick Wooten}
\address{Department of Applied Science, University of California at Davis,
Davis, California 95616}

\maketitle

\begin{abstract}

Numerical studies of amorphous silicon show that the lowest
4\% of vibrational modes are plane-wave like (``propagons'') and the
highest 3\% of modes are localized (``locons'').  
The rest are neither plane-wave like nor localized.  We call
them ``diffusons.''  Since diffusons are by far the most
numerous, we try to characterize them by calculating 
such properties as wavevector and polarization (which seem
not to be useful), 
``phase quotient'' (a measure of the
change of vibrational phase on between first neighbor atoms),
spatial polarization memory, and diffusivity.  Localized states
are characterized by finding decay lengths, inverse participation
ratios, and coordination numbers of the atoms participating.

\end{abstract}
\pacs{61.43.Dq, 63.50.+x, 66.70.+f}


\section{Introduction}

Vibrational properties of disordered media have been reviewed
by various authors, in particular Elliott and Leath (1975),
Weaire and Taylor (1980), Visscher and Gubernaitis (1980),
and Pohl (1998).  Here we review and present new results in
a program of numerical study of vibrations of amorphous Si.
Among the new results not contained in earlier reviews are
theoretical treatments of heat conductivity and thermalization
rates in glasses.

Harmonic normal modes of vibration can be rigorously classified
as extended (E) or localized (L).  In three dimensions the vibrational
spectrum has sharp E/L boundaries (``mobility edges'')
separating these two kinds of modes.

There is another boundary, not sharp, which Mott called the
``Ioffe-Regel limit'' and which we call the 
``Ioffe-Regel crossover.''  This P/D boundary separates the spectrum 
into a region with ballistic propagation (P) where wavevector is a 
reasonably good quantum number and a region with only diffusion (D) 
where wavevector cannot be defined but states are still extended.
In region P, wave-packets
can travel at sound velocity over distances of at least two or
three interatomic spacings before scattering from disorder
(Allen and Kellner, 1998).  The distance
of ballistic propagation is the mean free-path $\ell$.  In
region D, only diffusive propagation occurs over any meaningful
distance, and the concepts of mean free path and wavevector lose
usefulness.  Although it may seem natural that the Ioffe-Regel
crossover should be close to the mobility edge (Alexander, 1989), it is
not true for the models we have studied.  Indeed, Mott and Davis (1971)
emphasize the non-coincidence.  Both regions P and D lie in the extended
(E) part of the spectrum.

\begin{figure}[t]
\centerline{\psfig{figure=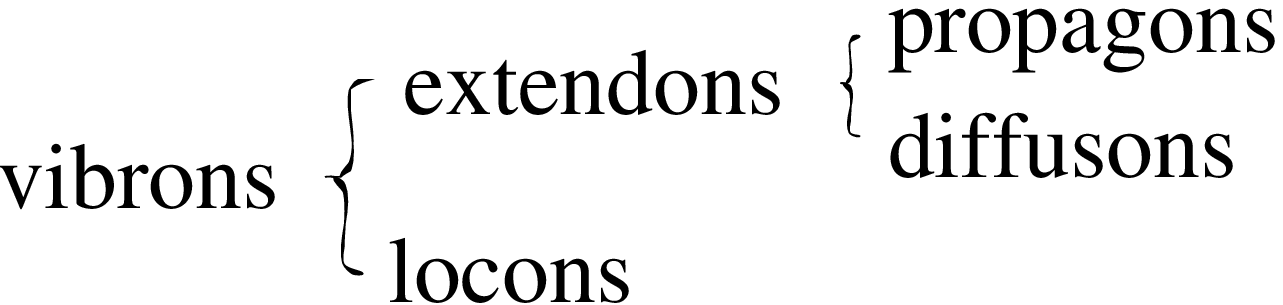,height=0.7in,width=2.7in,angle=0}}
\caption{Taxonomy of vibrations in glasses }
\label{zoology}
\end{figure}

The non-coincidence of the E/L boundary and the P/D boundary
means that the spectrum has three kinds of states.  We find
useful the terminology ``propagon, diffuson, and locon''
given in Fig. \ref{zoology}.  The term ``phonon'' is avoided because
of disagreement about what it means in a glass.

\par
\begin{figure}[t]
\centerline{\psfig{figure=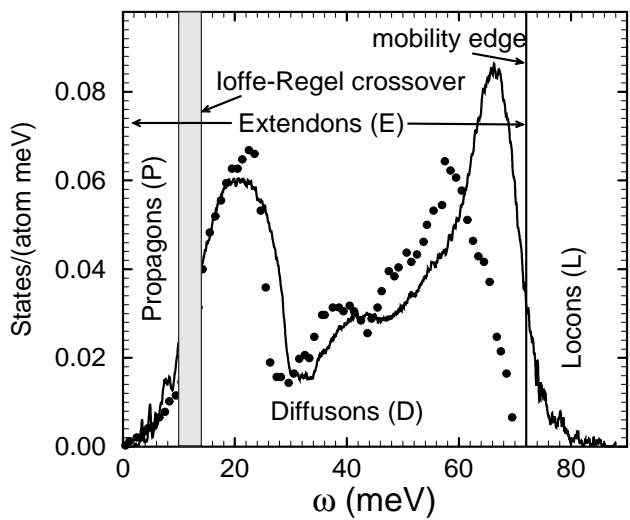,height=3.0in,width=3.5in,angle=0}}
\caption{Density of vibrational states from the 4096-atom model
compared with data from Kamitakahara \protect{\sl et al.} (1987).}
\label{dos}
\end{figure}

\par

We instinctively seek additional labels, to replace the 
detailed classification scheme by wavevector and branch so
useful in crystals.  Here we attempt to characterize as
completely as possible the dominant diffuson portion of
the spectrum, but fail to find useful sub-categories.
As already noticed by Kamitakahara {\sl et al.} (1987),
all diffuson modes of a given frequency have essentially
indistinguishable displacement patterns.

We study amorphous silicon in harmonic approximation,
using a model we believe to be very realistic.  As shown in Fig. \ref{dos},
the E/L boundary is near the top of the spectrum, with only
3\% of the modes localized.  The P/D boundary is near the bottom
of the spectrum, with only 4\% of the modes ballistically
propagating.  Diffusons fill 93\% of the
spectrum.  We do not think this is special either to our
model or to amorphous Si.  Other models for
amorphous Si (Kamitakahara {\sl et al.} 1987,
Lee {\sl et al.}, 1991, Nakhmanson and Drabold, 1998)
agree that the E/L boundary occurs near the top of the spectrum.
Similar results are found for other glasses 
(Bouchard {\sl et al.} 1988; Feldman and Kluge, 1995; 
Cobb, Drabold, and Cappelletti, 1996;
Taraskin and Elliott, 1997; Carles {\sl et al.}, 1998), 
and model systems (Sheng and Zhou, 1991; Sheng, Zhou, and Zhang, 1994;
Schirmacher, Diezemann, and Ganter, 1998), with
localized states occurring only near the top of the spectrum
or in tails near gaps in the vibrational densities of states.
For 3-d systems with artificially large disorder, the mobility
edge can be moved down to the middle of the spectrum
(Canisius and van Hemmen, 1985; Fabian and Allen, 1996).
The position of the P/D boundary near the bottom of the
spectrum is widely accepted.

\section{Low-frequency anomalies}

Our study of normal modes by
exact diagonalization on finite-size systems inherently
lacks information at low frequency.  
For this reason, the present paper ignores the well-known
but only partially understood low-frequency anomalies in
vibrational properties of glasses.  Our previous work
(Feldman, Allen, and Bickham, 1999) argues that the homogeneous
network models we use probably contain no low-frequency anomalies.
For completeness, we give here a brief catalog and point to 
sources for further information.

``Two-level'', or ``tunneling'' systems were introduced by
Phillips (1972) and Anderson, Halperin, and Varma (1972) motivated
by experimental discoveries by Zeller and Pohl (1971).
The predictive strength of this concept is beyond question, but the
physical objects the theory invokes remain elusive.
A review was given by Phillips (1987). 

The phrase ``boson peak'' refers to a low frequency feature seen
by Raman scattering (Stolen, 1970; J\"ackle, 1981) in many glasses, 
which is correlated with the
occurrence of ``excess modes'' in specific heat and other spectroscopies.
There are many candidate explanations for these modes.
One unified view, introduced by Karpov, Klinger, and Ignat'ev (1983) is called 
the ``soft potential model.''
This model holds that glasses generically have anharmonic regions, modelled as
double-well potentials.  These give rise to 
two-level systems , relaxational behavior, and quasi-localized or resonant
harmonic normal modes.  The last are a logical candidate for the
excess modes.  The subject was reviewed by Parshin (1994).
Our models contain some quasi-localized modes at low frequencies, 
as is mentioned further in Sec. VIII.

``Floppy modes'' (Phillips, 1980; Thorpe, 1983) and the daughter concept of
``rigid unit modes'' (RUMs: Dove {\sl et al.} 1996; Trachenko
{\sl et al.} 1998) refer to low-frequency modes which have zero
restoring force in nearest-neighbor central-force models.
Constraint counting algorithms provide methods of estimating
the numbers of such modes.  They are expected to be quasi-localized
in harmonic approximation, but intrinsically highly harmonic.
Such modes probably do not play any important role in amorphous
Si because of the overconstrained coordination.

Finally, experimental evidence shows that amorphous Si contains
usually fewer two-level-type excitations than most glasses, and that
samples with essentially zero such excitations can be prepared by
treatment with hydrogen (Liu {\sl et al.} 1997).

\section{The model}

Amorphous Si is an over-constrained network glass.
By one usual definition, it is not a glass
since it is not produced by a glass transition upon cooling.
Instead, thin films are condensed on cold substrates.  When
the film gets too thick, crystallization cannot be prevented.
The absence of a glass transition can be attributed
to good kinetic properties of a single-component
system with one strongly-preferred bonding arrangement.
To our mind, this just means that amorphous Si is simpler
than most glasses, which for our purposes is more of an advantage
than a disadvantage.  With obvious caution required, we think
that most of the properties we shall discuss can be regarded as
typical of most glasses.

\par
\begin{figure}[t]
\centerline{\psfig{figure=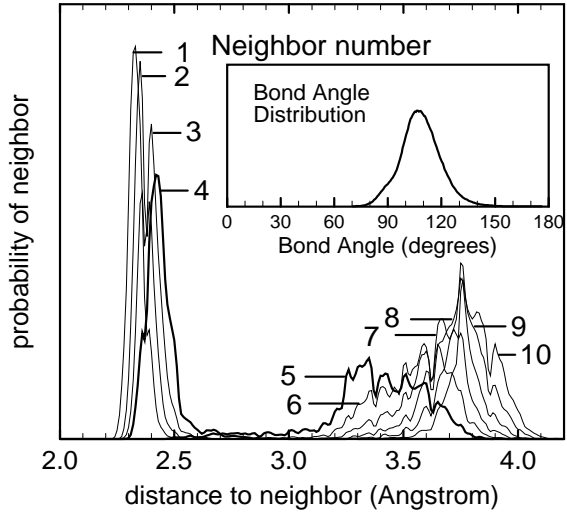,height=3.0in,width=3.5in,angle=0}}
\caption{Distribution of $n$th neighbors in the 4096-atom model of
amorphous silicon.  Fourth and fifth neighbors are shown in bolder
lines.  Note the small number of distant fourth or close fifth neighbors.
The inset shows the distribution of bond angles averaged over
all pairs of bonds with a common center and distances less than 2.6 $\AA$.}
\label{neigh}
\end{figure}

For amorphous Si we use atomic coordinates generated 
using the algorithm of Wooten, Weiner, and Weaire (1985).
We have studied models containing 216, 1000, and 4096 atoms,
contained in cubic boxes of side 16.5, 28, and 44$\AA$, respectively, 
and continued periodically to infinity.  All models are built using
the Keating potential (Keating, 1966) and then subsequently re-relaxed
to a local minimum of the Stillinger-Weber potential (Stillinger and
Weber, 1985).  Different models differ in
minor details, both for ordinary statistical reasons and because the
algorithm was implemented slightly differently in each case.
In this section we present structural properties of a 4096-atom model.
This model has larger distortions from tetrahedral form than some
of our other models, with 102 four-membered rings.
Figure \ref{neigh} shows neighbor distributions ${\cal P}_n(\omega)$
The distribution ${\cal P}_1$ for the first
neighbor is calculated by measuring the distance to the closest neighbor
of each atom; ${\cal P}_2$ is found from the distances to the second closest
neighbor.  Note that the first four neighbors distribute tightly
within $\approx 0.1\AA$ to the crystalline first neighbor distance,
2.35$\AA$.  The next 12 neighbors in crystalline Si are at 3.84$\AA$, 
a number fixed by the bond length (2.35$\AA$) and the tetrahedral
bond angle (109.5$^{\circ}$).  In
our model of amorphous Si, the closest of the next 12
(neighbor number 5) lies roughly between 3.2 and 3.7$\AA$,
while the farthest of these 12 (neighbor number 16, not shown) lies roughly
between 3.9 and 4.4$\AA$.  This reflects a flexibility in the bond
angles, with values distributed between 90$^{\circ}$ and 125$^{\circ}$,
as shown in the inset to Fig. \ref{neigh}.  
The third set of neighbors in crystalline
Si is 12 atoms at 4.50$\AA$.  This is determined by the fact that
diamond structure has a dihedral angle of 60$^{\circ}$, with
all rings of the 6-member ``chair'' type.  Rotation of the dihedral
angle to 0$^{\circ}$ (``boat'' type rings) reduces the third neighbor
distance to 3.92$\AA$.  Our model shows no gap at all between
second shell (neighbors 5-16) and third shell (neighbors 17-28),
consistent with random dihedral angles.

\begin{figure}[t]
\centerline{\psfig{figure=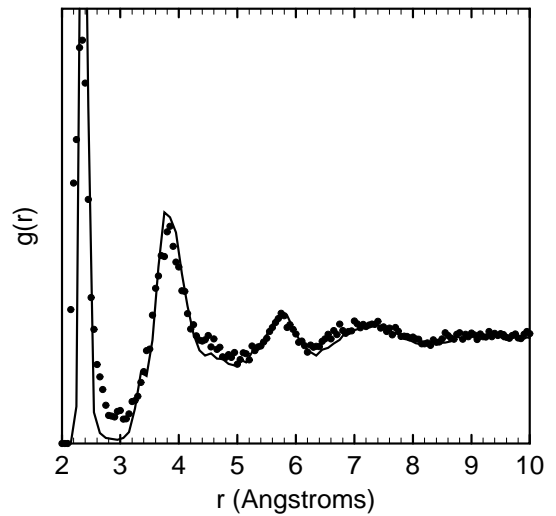,height=3.0in,width=3.5in,angle=0}}
\caption{Radial distribution function obtained by summing the
$n$th neighbor distributions of Fig. \protect\ref{neigh} over all $n$,
and dividing out a factor $r^2$. The data are from a neutron diffraction
experiment by Kugler \protect{\sl et al.} (1993). }
\label{gofr}
\end{figure}

The sum $\sum {\cal P}_n$ over all $n$ gives the radial distribution
function $g(r)$, plotted in Fig. \ref{gofr} and compared with the experiments
of Kugler {\sl et al.}  The close agreement is one measure of the
realism of the model.  However, much of the structure of $g(r)$ seems
only to reflect atom density and nearest neighbor distance, so it
may not be a very stringent test.

\par

\section{Vibrational frequencies}

The other aspect of the model is the interatomic forces.  
We chose the model of Stillinger and Weber (1985) which is designed to work
for the liquid as well as crystalline state.  This required us to
relax the coordinates from Wooten to a minimum of the Stillinger-Weber
potential.  The stability of the minimum is proven
by the positivity of all eigenvalues $\omega^2$ of the dynamical
matrix.  The eigenfrequency distribution is shown in Fig. \ref{dos}.
Qualitatively satisfactory agreement is found with the neutron scattering
data of Kamitakara {\sl et al.} (1987).  A similar overestimate
of vibrational frequencies is made when Stillinger-Weber forces
are applied to crystalline Si, so we think the discrepancies
should be attributed to the forces rather than the atomic coordinates.

\section{localized states}

\begin{figure}[t]
\centerline{\psfig{figure=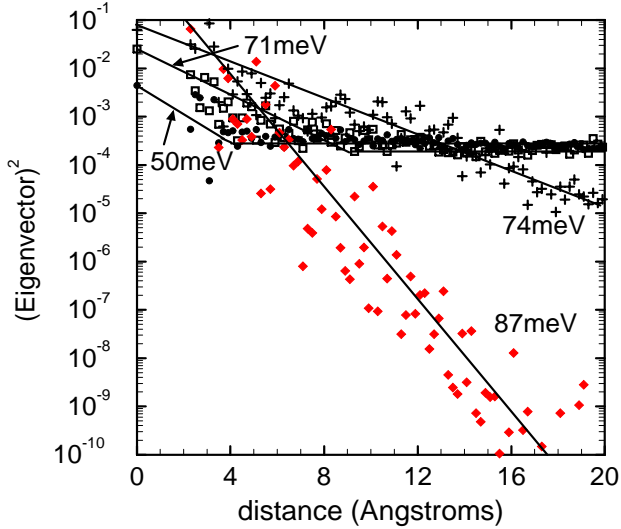,height=3.0in,width=3.5in,angle=0}}
\caption{Spatial decay of vibrational eigenvectors $\epsilon_i$.
Solid lines are guides to the eye.
For each mode $i$ the atom with the largest $|\epsilon|^2$ is
located, and taken as the origin.  The plot shows values of
\protect$|\epsilon(\vec{R}_n)|^2$ averaged over spherical shells of
\protect$|\vec{R}_n|$ of width $\delta R=0.2\AA$.
Results shown for $\omega$=50 meV, 71 meV, 74 meV, and 87 meV, are
averages over 3, 4, 2, and 2 modes respectively.  Modes at 50 meV
and 71 meV have mean square eigenvector near 1/($N$=4096)
throughout the cubic cell of length 44 $\AA$, with enhanced values only
in spatial regions of small measure.  These modes are extended, as are all the
modes below the mobility edge at 73 meV.  Modes at 74 meV and 87 meV
have mean values of \protect$|\epsilon(\vec{R}_n)|^2 \approx
\protect\exp(-2|\vec{R}_n|/\xi)$ with localization length $\xi=4.7\AA$ and
1.5$\AA$ respectively.}
\label{eigfall}
\end{figure}

The definition of a localized state is exponential decay of
the eigenvector with distance from some center $\vec{R}_0$:
\begin{equation}
|\vec{\epsilon}_i(\vec{R}_n)|\propto \exp(-|\vec{R}_n -\vec{R}_0|/\xi_i).
\label{eq:xi}
\end{equation}
This defines the localization length $\xi_i$ of the $i$-th normal
mode, if the decay is observed.  Fig. \ref{eigfall} shows 
selected modes, showing the very different character of modes from
the D and the L portions of the spectrum.

\par

There is still controversy concerning the location
of the mobility edge in glasses.   Unfortunately, most 
experiments shed little light, since measured spectral
properties, being averages over a macroscopic region, do not
differentiate between localized and delocalized states.
Heat conductivity $\kappa(T)$ is the property most strongly affected
by localization.  We think the measured $\kappa(T)$ strongly
supports our placement of the P/D and E/L boundaries
near the lower and upper edges of the spectrum.  This
will be discussed in the next section. 

\begin{figure}[t]
\centerline{\psfig{figure=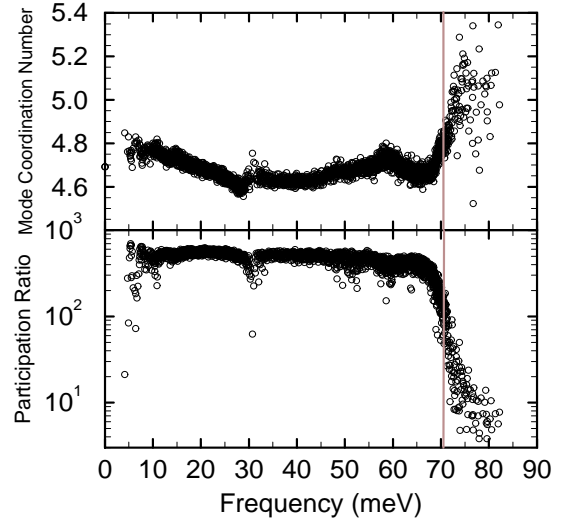,height=3.0in,width=3.5in,angle=0}}
\caption{The mode average coordination number (Eqn. \protect\ref{eq:zi}),
and the participation ratio (on a logarithmic scale), 
calculated for a model with 1000 atoms.}
\label{mocopr}
\end{figure}

If eigenvectors are calculated, then
no special tricks are needed to locate the theoretical
mobility edge in models of amorphous silicon.   
A model with 216 atoms was large enough to locate
the E/L boundary between 71 and 73 meV; the
precise location varies somewhat from model to model.
We have made some experiments with artificially enhanced disorder,
randomly scaling half of the masses by a factor of 5.
This pushes the E/L boundary into the middle of the spectrum 
where it becomes more blurred by size effects.
Localized states of ``pure'' amorphous Si are easily recognized.
They are trapped in especially defective regions.  This was discovered
(Fabian, 1997b) by defining a local
coordination number $Z_a$, the number of neighbors
of atom $a$ at $\vec{R}_a$.  The following arbitrary
definition of neighbor suffices:
\begin{equation}
Z_a = \sum_b w(|\vec{R}_a-\vec{R}_b|)
\label{eq:za}
\end{equation}
where $w(r)$ is 1 for $r<2.35\AA$, 0 for $r>3.84\AA$, and continuous
and linear in between.  This gives an average coordination of 4.7
neighbors.  The ``mode average coordination number'' is defined as
\begin{equation}
Z_i = \sum_a Z_a \left|\vec{\epsilon}_i (a)\right|^2.
\label{eq:zi}
\end{equation}
Most modes have $Z_i$ near average, but localized modes mostly
collect at regions with significantly higher coordination, 
as shown on Fig. \ref{mocopr}.

\par

We have seen the mobility edge in many independent calculations.
\begin{enumerate}
\item The participation ratio $p_i$ (Bell and Dean, 1970),
plotted on Fig. \ref{mocopr},
gives the number of atoms on which the mode has significant amplitude.
For the E=P+D part of the spectrum, the value hovers near 500 for a
model with 1000 atoms.
\item The diffusivity $D_i$ drops to
zero at the mobility edge.  This is shown in the next section.
\item Sensitivity to boundary conditions is also discussed
in the next section.
\item Level spacing distributions have been computed 
for diffusons and locons (Fabian, 1997b).  As expected,
diffusons obey Wigner-Dyson statistics, while locons obey 
Poisson statistics.
\item Eigenvector self-correlation functions were carefully
studied in a 4096-atom model by Feldman {\sl et al.} (1998).
\item Many quantities $q_i$ which can be evaluated for each
mode $i$ seem to depend only on $\omega_i$ for diffusons,
but become mode-specific for locons.  The ones we have looked at
are:
\begin{description}
\item{(a.)}
The mode-average coordination number, plotted on Fig. \ref{mocopr}. 
\item{(b.)} The phase-quotient parameter, discussed in section VII.C.
\item{(c.)} Bond-stretching
parameters (Fabian and Allen, 1997).
\item{(d.)}
Gr\"uneisen parameters for both volume and shear deformations
(Fabian and Allen, 1997 and 1999).
\end{description}
\end{enumerate}

\section{heat conductivity and diffusivity}

Theoretical interpretation of heat conduction of glasses
has been contentious.
Below the ``plateau'' region (typically 5-30K) it is agreed that heat
is carried by ballistically propagating low-frequency modes (the P
region of the spectrum).  Above the plateau, $\kappa(T)$ rises,
approaching at room temperature a constant value which is typically
smaller than the crystalline value (a decreasing function of
$T$ at room temperature.)  A rigorous consequence (Allen and
Feldman, 1993) of the Kubo
formula (Kubo, 1957) and the harmonic approximation is the relation
\begin{equation}
\kappa(T)=\frac{1}{V}\sum_i C(\hbar\omega_i/2k_B T) D_i,
\label{eq:kappa}
\end{equation}
where C(x) is the specific heat of a harmonic oscillator
$(x/\sinh(x))^2$
and $D_i$ is the ``diffusivity'' of the
$i$-th normal mode of frequency $\omega_i$, given by
\begin{equation}
D_i=\frac{\pi V^2}{3\hbar^2\omega_i^2}\sum_{j\ne i}|S_{ij}|^2
     \delta(\omega_i -\omega_j),
\label{eq:diff}
\end{equation}
where $S_{ij}=<i|S|j>$ is the intermode matrix element of the
heat current operator.
Eq. \ref{eq:kappa} also emerges, with $D_Q$ equal to $v_Q^2 \tau_Q /3$,
from the Peierls-Boltzmann phonon-gas
model (Gurevich, 1986) of transport in crystals.  The latter model is only
justified if
the mean free-path $\ell_Q=v_Q \tau_Q$ is longer than a wavelength.

It was noticed by Birch and Clark (1940), and by Kittel (1948) 
that in glasses $\kappa(T)$  at $T>$20K could be interpreted
as the specific heat $C(T)/V$ multiplied by a temperature-independent
diffusivity $\bar{D}$ of order $a^2 \omega_D /3$
where $a$ is an interatomic distance.   In the phonon-gas model,
this would correspond to $\ell\approx a$, too small to justify
use of the model.  The success of this observation implies
that the dominant normal modes in a glass are of the D variety, not
P because P implies $\ell\gg a$, and not L because L implies $D=0$
until anharmonic corrections are added which make $D$ depend on $T$.
This successful
(and we believe, essentially correct) interpretation lost
favor after Anderson localization was understood, because a
misconception arose that the P/D boundary (which certainly lies
low in the spectrum of a glass) should lie close to the
E/L boundary.  

\begin{figure}[t]
\centerline{\psfig{figure=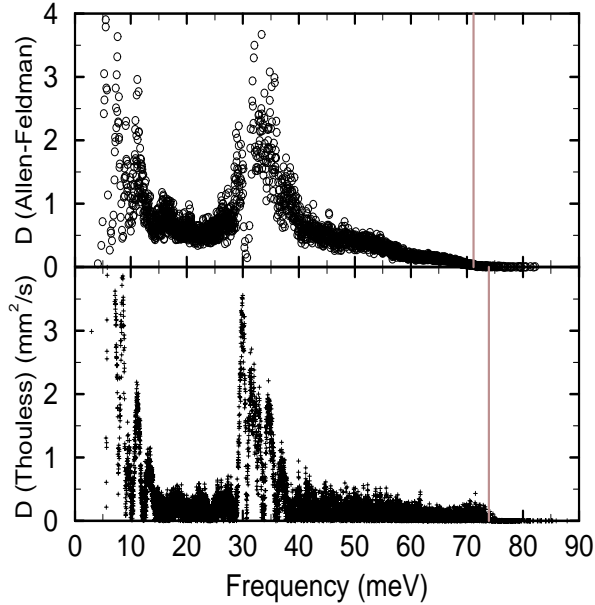,height=3.5in,width=3.5in,angle=0}}
\caption{Diffusivity versus frequency
(Feldman \protect{\sl et al.}, 1993, Fabian, 1997b) calculated
for a 1000 atom model by Eq. \protect\ref{eq:diff}, and for a
4096 atom model by Eq. \protect\ref{eq:thouless}.  Mobility edges,
shown as vertical lines, are slightly different in these two models. }
\label{diff}
\end{figure}

Our numerical calculations of $D_i$ are shown in Fig. \ref{diff}.
Also shown are values of $D_i$ from a formula of Edwards and Thouless
(1972),
\begin{equation}
D_i=L^2 \Delta\omega_i,
\label{eq:thouless}
\end{equation}
where $\Delta\omega_i$ is the sensitivity of the eigenenergy
to a twist of the boundary condition.  We have simply used
for $\Delta\omega_i$ the change in $\omega_i$ when boundary 
conditions are changed from periodic to antiperiodic.  The
actual definition is probably
\begin{equation}
\Delta\omega_i = \lim_{\phi\rightarrow 0}\left[ \frac{\pi^2}{\phi^2}
   \Delta\omega_i(\phi)\right]
\label{eq:shift}
\end{equation}
where $\Delta\omega_i(\phi)$ is the shift when the phase is twisted
by $\phi$.  Antiperiodic boundary conditions correspond to
$\phi=\pi$, while $\phi=2\pi$ returns to periodic boundary
conditions with $\Delta\omega_i(2\pi)=0$.  Therefore our calculation,
which is the only one easily accessible for us, gives a probable
upper bound to $D_i$ for each mode $i$.  Inspection of Fig.
\ref{diff} shows that with this interpretation, the two
calculations agree reasonably well.  Both go to zero at the
mobility edge, and both become large and ragged in the P region
below 10 meV.  The raggedness comes from the sparseness of the
eigenstates at low $\omega$, and the large values reflect the
onset of ballistic propagation.  In the D region above 12 meV,
values have collapsed to the range of 1 mm$^2$/s, which corresponds
to $\omega_D a^2 /3$, with $\omega_D$ set to 50 meV and $a=2\AA$.
This diffusivity is well below any value that could be allowed in
a phonon-gas picture, and agrees with the measured $\kappa(T)$
(Allen and Feldman, 1990; Feldman {\sl et al.}, 1993).
The peak of $D_i$ around 33 meV corresponds to a feature in the
``phase-quotient'' that will be discussed in section VII.C.
Similar results for vitreous SiO$_2$ have been reported by
Feldman and Kluge (1995).

\par

\begin{figure}[t]
\centerline{\psfig{figure=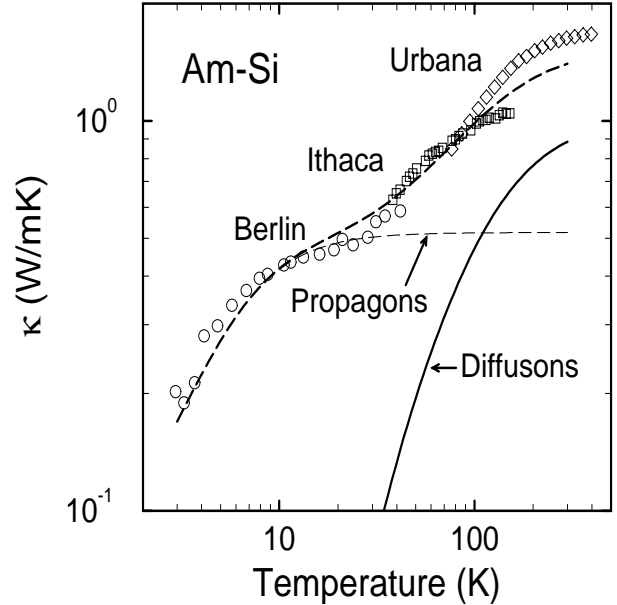,height=3.5in,width=3.5in,angle=0}}
\caption{Thermal conductivity $\kappa$ versus temperature $T$
measured for amorphous silicon (circles: Pompe and Hegenbarth, 1988;
squares: Cahill \protect{\sl et al.}, 1989; diamonds: Cahill 
\protect{\sl et al.}, 1994).  The bold line is a calculation
from the diffusivity values shown in Fig. \protect\ref{diff}
for a 1000 atom model using Eq. \protect\ref{eq:kappa}. 
The thin dashed line is a calculation from modes at low energy assuming
that their diffusivity obeys an $\omega^{-2}$ law.  The bold dashed
line is the sum of these two contributions.}
\label{kappa}
\end{figure}

When $\kappa(T)$ is calculated from Eq. \ref{eq:kappa}, using
the values of $D_i$ from Fig. \ref{diff}, the results, shown in
Fig. \ref{kappa} agree roughly with the data at higher temperatures.
At low temperature it is necessary to have an additional source
of heat current, the ballistically propagating long-wavelength
modes.  In Fig. \ref{kappa}, this has been added in a thoroughly
{\sl ad hoc} fashion.  We have simply assumed a Debye spectrum
for the modes with $\omega<\omega_0=$3 meV, and a temperature independent
diffusivity $D(\omega)=D_0 (\omega_0/\omega)^2$.  There is no theoretical
justification for this.  In principle, the temperature-independent
diffusivity caused by glassy disorder should take the Rayleigh
$(\omega_0/\omega)^4$ form at low $\omega$, and one needs a stronger
type of scattering, inelastic, and therefore $T$-dependent, to 
match the data.  However, the $(\omega_0/\omega)^2$ behavior has
been seen at intermediate frequencies, both experimentally
(Sette {\sl et al.} 1998)
and numerically (Dell'Anna {\sl et al.} 1998; Feldman {\sl et al.} 1999),
so we have used this simpler fitting proceedure
to match the data.  

\par

Our most important conclusion is that the ``re-increase'' of thermal
conductivity above the plateau region is attributable to heat
carried by ``diffuson'' modes in much the way imagined by Birch, Clark,
and Kittel, and that the plateau is a simple crossover region, not
requiring any new physics to explain.  In particular, we believe
that ``excess modes'' (also known as a ``Boson peak'') is not a
necessary ingredient to explain the plateau.  Amorphous silicon
seems to lack these ``excess modes'' but still to have a plateau.

\section{thermal equilibration}

There is some evidence suggesting that vibrations in glasses, if 
disturbed from equilibrium, may return very slowly.  For amorphous
silicon, experiments were reported by Scholten and 
Dijkhuis (1996) and by Scholten, Akimov, and Dijkhuis (1996).
Our investigations
show that if the disturbance is not too large and is purely vibrational,
then the rate of return to equilibrium should be as fast, if not faster,
than in a corresponding crystal.  Surprisingly, we find that this is
true both in the locon and in the diffuson portion of the spectrum,
contradicting a view (Orbach, 1996) supported by fracton theory
(Alexander, 1989) that localized vibrations must equilibrate slowly.

\par
\begin{figure}[t]
\centerline{\psfig{figure=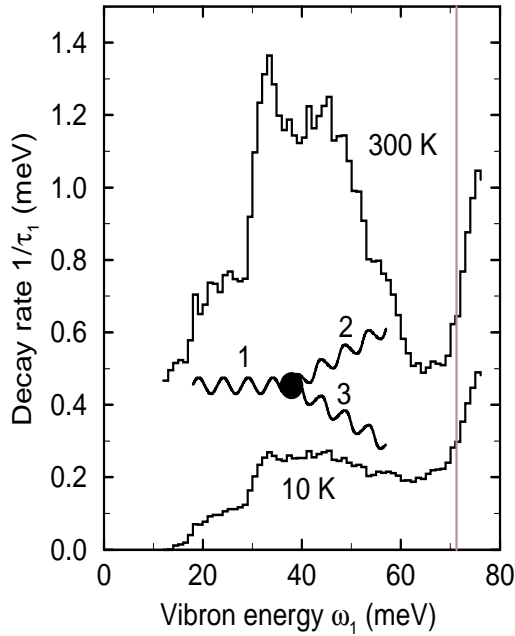,height=3.5in,width=3.5in,angle=0}}
\caption{Decay rates calculated by anharmonic perturbation
theory for amorphous Si.  The inset shows the process diagrammatically.
}
\label{decay}
\end{figure}

In thermal equilibrium, the harmonic vibrational eigenstates
have average population given by the Bose-Einstein distribution.
Fabian and Allen (1996) used anharmonic perturbation theory
to calculate the inverse lifetime or equilibration rate $1/\tau$
by which a vibrational state returns to the Bose-Einstein
distribution if driven out of equilibrium.  Their results are shown
in Fig. \ref{decay}.  The validity of the
perturbation theory is confirmed both by the smallness of
the ratio $1/\omega\tau$ and by a direct test in the
classical regime using molecular dynamics 
by Bickham and Feldman (1998) and Bickham (1999) 

It can be seen in Fig. \ref{decay} that no change occurs
in the size of $1/\tau$ at the mobility edge.
A careful treatment of locons does
not support the idea that they thermalize slowly.  Their
anharmonic thermalization rates (Fabian, 1997a)
are as fast, or even faster, than
diffusons, and also comparable to, or faster than, corresponding
thermalization rates of vibrations in crystals.  
The source of the misconception of slow equilibration is the
idea that direct decay of a locon into two locons should be
negligible.  This idea fails because, unlike for example, 
band-tail electronic states in amorphous Si, the vibrational
states are not at all dilute.
Slow thermalization rates (forbidden by all theories we
understand) could be tested by
looking for a contribution from thermal vibrations to attenuation of
very high-frequency sound (Fabian and Allen, 1999).  If the
thermalization rate is extremely slow, this contribution
to the attenuation would be greatly enhanced.

\section{resonant modes}

Inspection of Fig. \ref{mocopr} shows that a few modes in the D
region and somewhat more in the P region have anomalously small
participation ratios, of order 100 out of the 1000 atoms available.
These states are not exponentially localized (Fabian, 1997b; Feldman
{\sl et al.}, 1999) but are temporarily trapped in regions of
peculiar coordination, from which they can tunnel into the 
continuum of extended states.  Such states were first reported by
Biswas {\sl et al.} (1988) in a small model similar to ours;
a model with larger numbers of 3- and 5-fold coordinated atoms
had more such modes, which were speculated to bear some relation
to the ``floppy modes'' of Phillips and Thorpe.
Such modes were studied in detail by Schober and coworkers (1988, 1991)
and Oligschleger and Schober (1999).

We have recently argued (Feldman {\sl et al.}, 1999) 
that, in our (mostly 4-fold coordinated)
models of amorphous Si, such states tend to disappear
as the size of the model gets bigger, presumably because each
such mode is trapped only in a very specific peculiar region.
As the number of atoms in the model increases, so does the
number of peculiar regions, but if each resonant mode is trapped
in only one region, the fraction of time spent outside that region
increases because the volume outside that region has increased.  
On the other hand, such modes, especially the ones in the P region,
may be more pronounced in real amorphous Si and other real glasses
than they are in the models we study.  This is because our models
are spatially homogeneous on scales greater than $4\AA$, while
real glasses may have mesoscopic defects such as voids which would
attract more such modes.

Fabian and Allen (1997, 1999) found that the resonant modes have
giant ($\approx -40$) Gr\"uneisen parameters $\gamma_i$.
These parameters measure the sensitivity of $\omega_i$ 
to macroscopic strain.  In a glass (just as in a crystal where
positions of atoms are not all fixed by crystallographic
symmetry) strains cause not just a homogeneous shift of atomic
coordinates, but also an additional local relaxation, which turns
out to be particularly large in just those peculiar regions
where the resonant modes are temporarily trapped.
Anomalously large values of $\gamma_i$ play an important role in
explaining the anomalously large and sample-dependent measured
low-$T$ thermal expansion of glasses, and also should show up in
enhanced contributions to the attenuation of high $\omega$ sound waves
at higher $T$.

\section{character of diffusons}

The most important property which distinguishes diffusons is
their intrinsic diffusivity $D_i$ with values of order $\omega_D a^2 /3$.
If wave-packets were constructed in such a way as to be approximately
monochromatic, and simultaneously localized at the center
of a cell with a reasonably
small radius (perhaps 6-8$\AA$), then we believe that no matter how
well directed such a pulse was at $t=0$, the center of the pulse
would hardly move, and the radius would evolve as $<r^2>=6Dt$
for all times until reaching the cell boundary where it would
interfere with its periodic image.  Unfortunately, a 44$\AA$
cell is only marginally big enough, and computational difficulties
have so far prevented us performing this experiment.
Here we describe our efforts to find other ways to characterize diffusons.

\subsection{wavevector}

\begin{figure}[t]
\centerline{\psfig{figure=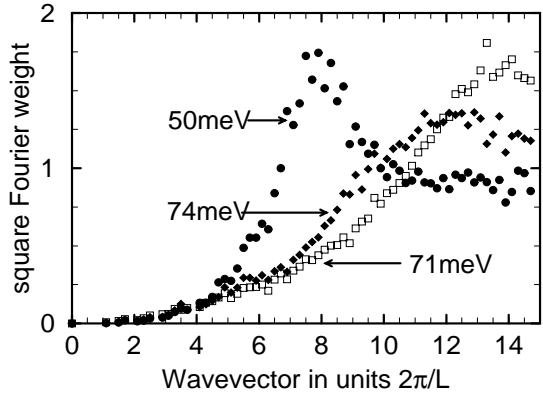,height=3.0in,width=3.5in,angle=0}}
\caption{Fourier weights calculated from Eq. \protect\ref{eq:fourier}
for diffusons at 50 meV and 71 meV, and locons at 74 meV, averaged
over 3, 4, and 2 modes respectively. }
\label{fourier}
\end{figure}

At the P/D boundary, wavevectors become ill-defined.  Fig. \ref{fourier}
shows a test.  The squared Fourier weight is defined as
\begin{equation}
w_i(\vec{Q})=\left| \sum_{\ell} e^{i\vec{Q}\cdot\vec{R}(\ell)}
\vec{\epsilon}_i(\vec{R}(\ell)) \right|^2
\label{eq:fourier}
\end{equation}
where $\vec{Q}$ is chosen as $(2\pi/L)(h,k,l)$ with integer $h,k,l$
so that the periodic images interfere constructively.  We define
$w_i(Q)$ as $w_i(\vec{Q})$
averaged over spherical shells of wave vector of width
$0.2\times 2\pi/L$.  The 
value $Q=9.2\times 2\pi/L$ corresponds to neighboring
atoms being completely out of phase.  The 51 meV diffusons show
a peak Fourier content near $8\times 2\pi/L$, but the peak height is less than
twice a ``background'' value found at larger $Q$ which dominates
the behavior.  There is no ballistic character to these modes.

\par

\subsection{polarization}

\begin{figure}[t]
\centerline{\psfig{figure=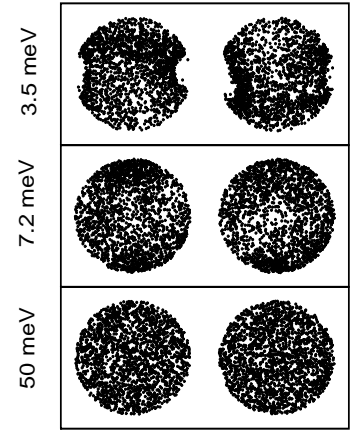,height=2.5in,width=3.5in,angle=0}}
\caption{Area-preserving projection normalized unit polarizations 
$\hat{\epsilon}_i(\vec{R}(\ell))$ onto circles (one circle for
each hemisphere.)  Three modes $i$ are shown, with each of the 4096
components showing as a dot. }
\label{polar}
\end{figure}

Diffusons have no wave vector.   Not surprisingly, they also
lack a polarization as is shown in Fig. \ref{polar}.  
Propagons, by definition, have a wavevector.  The nature of
the propagons in the 4096 atom model was examined by
Feldman {\sl et al.} (1999).  As shown there,
the modes near $\omega=$3.5 meV have well-defined TA character, with
the smallest possible wave vector
$Q=2\pi/L$.  Fig \ref{polar} shows that these modes have only
limited preference for a direction of polarization.
Similarly, the mode at 7.2 meV has $Q=2\pi/L$ and
LA character, but not much polarization.

\par

\par
\begin{figure}[t]
\centerline{\psfig{figure=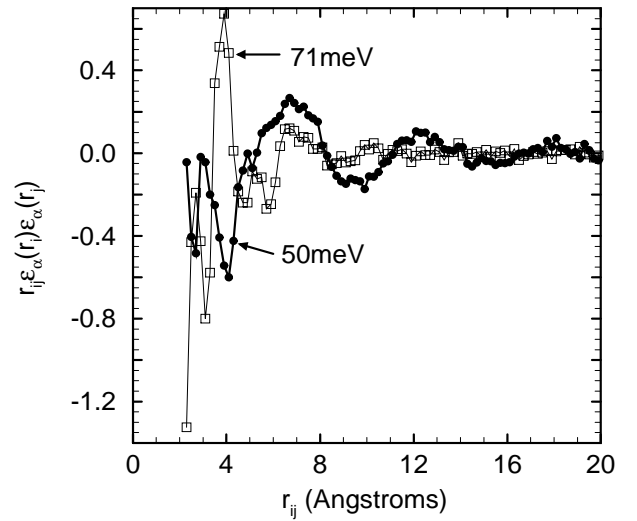,height=3.0in,width=3.5in,angle=0}}
\caption{The spatial fall-off of $\vec{\epsilon}(\vec{r}_i)
\cdot \vec{\epsilon}(\vec{r}_j)$ weighted by \protect$r_{ij}$ and
averaged for the same 3 modes
at 50meV and 4 modes at 71meV
as in fig. \protect\ref{eigfall}.  }
\label{eigcor}
\end{figure}

Polarization directions of diffusons wander uniformly over the unit
sphere, and one may ask what is the spatial range of decay of polarization
memory in a mode.  This is shown in Fig. \ref{eigcor}.
In a crystal, one has surfaces in $Q$-space where modes have constant
frequency $\omega(\vec{Q})=$const.  A single eigenstate is some
arbitrary linear combination of the degenerate Bloch waves on this
surface.  If the surfaces were spherical and the linear
combinations random, we would expect that $\vec{\epsilon}(\vec{r}_i)
\cdot \vec{\epsilon}(\vec{r}_j)$ would fall off spatially as
$\cos(Qr_{ij})/r_{ij}$.  Fig. \ref{eigcor} shows that for diffusons
at 50 meV, some polarization memory, but much less than expected
for a crystal, persists out to 12$\AA$, while higher $\omega$
diffusons lose polarization memory more rapidly.

\subsection{phase quotient}

\begin{figure}[t]
\centerline{\psfig{figure=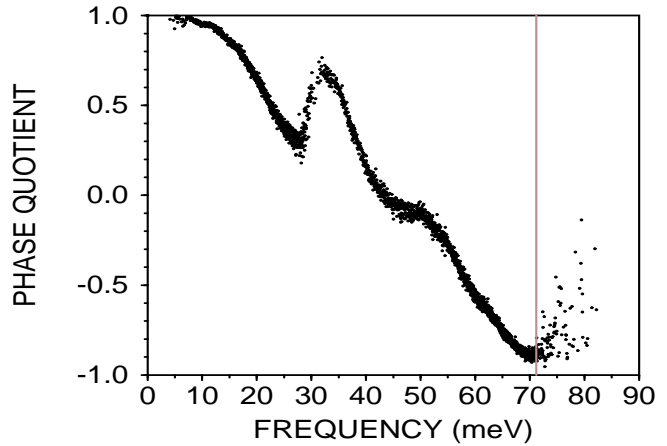,height=2.5in,width=3.5in,angle=0}}
\caption{Phase quotient versus energy for a 1000 atom model. }
\label{pq}
\end{figure}

The ``phase quotient'' $\phi_i$ was defined by Bell and
Hibbins-Butler (1975) as
\begin{equation}
\phi_i=\frac{\sum_{<a,b>}\vec{\epsilon}_i(a)\cdot\vec{\epsilon}_j(b)}
{\sum_{<a,b>}\left|\vec{\epsilon}_i(a)\cdot\vec{\epsilon}_j(b)\right|},
\label{eq:pq}
\end{equation}
and is plotted in Fig. \ref{pq}.  For low $\omega_i$, values near 1
indicate that nearest neighbor atoms (the only ones summed in
Eqn. \ref{eq:pq}) vibrate mostly in-phase, while for high $\omega_i$, 
values near -1 indicate that nearest neighbors vibrate mostly
out-of-phase.  Like so many other properties, this depends only
on $\omega_i$ until the E/L boundary is crossed, but is very
mode-specific for locons.  The sharp rise at $\omega_i \approx$ 29 meV
is interesting, and may help explain why at the same frequency
$D_i$ (Fig. \ref{diff}) has a sudden rise.  In crystalline Si,
at approximately the same point in the spectrum, the TA branch
ends and the density of states has a local minimum.  Thus 29 meV
marks a point in the spectrum where diffusons change character
from bond-bending (somewhat TA-like) with relatively high
frequency because of large phase difference from atom to atom,
to bond-stretching (somewhat LA-like) with not such a large
phase difference but an equally high frequency because the
bond-stretching forces are bigger than the bond-bending forces.
Apparently the latter kind of mode has greater diffusivity
by a factor 2 or more.  In a crystal we attribute this to
a larger group velocity of the LA branch and a smaller 
density of states for decay by elastic scattering.  Neither of
these properties can be properly invoked for diffuson modes
in a glass, but apparently similar physics is somehow at work.

\par

\section{summary}
Since 95\% of the states in amorphous Si, and probably many other
glasses as well, are diffusons, we should understand their properties.
All diffusons at a given frequency $\omega$ seem essentially identical.
As $\omega$ changes, their properties evolve, mostly smoothly,
but the sudden jump in diffusivity and in phase quotient at 29 meV
shows that not all variability is lost.  

\acknowledgements

We thank S. Bickham and
K. Soderquist for relaxing the coordinates,
and B. Davidson for providing eigenvalues and eigenvectors 
of the 4096 atom model. 
Work of PBA was supported in part by NSF grant no. DMR-9725037.
Work of JF was supported by the U. S. Office of Naval Research.

\begin{center}
{\bf REFERENCES}
\end{center}

\noindent		Alexander, S., Laermans, C., Orbach, R., and
			Rosenberg, H. M., 1983,
			Phys. Rev. B {\bf 28}, 4615;
			R. Orbach and A. Jagannathan, 1993,
			J. Phys. Chem. {\bf 98}, 7411.\\

\noindent		Alexander, S., 1989,
			Phys. Rev. B {\bf 40}, 7953.\\

\noindent		Allen, P. B., and Feldman, J. L., 1993,
        		Phys. Rev. B {\bf 48}, 12581.\\

\noindent		Allen, P. B., and Kelner, J., 1998
			Am. J. Phys. {\bf 66}, 497.\\

\noindent               Anderson, P. W., Halperin, B. I., and Varma, C. M., 
                        1972,
 			Philos. Mag. {\bf 25}, 1.

\noindent		Bell, R. J., and Dean, P., 1970,
                        Disc. Faraday Soc. {\bf 50}, 55.\\

\noindent		Bell, R. J., and Hibbins-Butler, D. C., 1975,
			J. Phys. C {\bf 8}, 787.\\

\noindent		Bickham, S. R.,  and Feldman, J. L., 1998,
			Phys. Rev. B {\bf 57}, 12234 and
			Phil. Mag. B {\bf 77}, 513.\\

\noindent		Bickham, S. R., 1999,
			Phys. Rev. B {\bf 59}, 4894.\\

\noindent		Birch, F., and Clark, H., 1940,
			Am. J. Sci. {\bf 238}, 529 and 612.\\

\noindent		Biswas, R., Bouchard, A. M., Kamitakahara, W. A.,
 			Grest, G. S., and Soukoulis, C. M., 1988,
 			Phys. Rev. Letters {\bf 60}, 2280.\\

\noindent		Bouchard, A. M., Biswas, R., Kamitakahara, W. A.,
			Grest, G. S., and Soukoulis, C. M., 1988,
			Phys. Rev. B {\bf 38}, 10499.\\

\noindent		Cahill, D. G., Fischer, H. E., Klitsner, T.,
			Swartz, E. T., and Pohl, R. O., 1989,
			J. Vac. Sci. Technol. A {\bf 7}, 1259.\\

\noindent		Cahill, D. G., Katiyar, M., and Abelson, J. R., 1994,
			Phys. Rev. B {\bf 50}, 6077.\\

\noindent		Canisius, J., and van Hemmen, J. L., 1985,
			J. Phys. C {\bf 18}, 4873.\\

\noindent		Carles, R., Zwick, A., Moura, C. and
			Djafari-Rouhani, M., 1998,
			Phil. Mag. {\bf 77}, 397.\\

\noindent		Cobb, M., Drabold, D. A., and Cappelletti, R. L., 1996,
			Phys. Rev. B {\bf 54}, 12162.\\

\noindent		Dell'Anna, R., Ruocco, G., Sampoli, M., and
			Viliani, G., 1998,
			Phys. Rev. Letters {\bf 80}, 1236.\\

\noindent		Dove, M. T., Giddy, A. P., Heine, V., and
			Winkler, B., 1996,
			Am. Mineral. {\bf 81}, 1057.\\

\noindent		Edwards, J. T., and Thouless, D. J., 1972,
			J. Phys. C. {\bf 5}, 807.\\

\noindent		Elliott, R. J., and Leath, P. L., 1975,
			in {\sl Dynamical Properties of Solids},
			(eds.  G. K. Horton and A. A. Maradudin),
			North Holland, Amsterdam; vol. 2, p.385.\\

\noindent		Fabian, J., and Allen, P. B., 1996,
        		Phys. Rev. Letters {\bf 77}, 3839.\\

\noindent		Fabian, J., 1997a,
			Phys. Rev. B {\bf 55}, R3328.\\

\noindent		Fabian, J., 1997b,
			Ph.D. dissertation, SUNY Stony Brook.\\

\noindent		Fabian, J., and Allen, P. B., 1997,
        		Phys. Rev. Letters {\bf 79}, 1885.\\

\noindent		Fabian, J., and Allen, P. B., 1999,
                        Phys. Rev. Letters {\bf 82}, 1478.\\

\noindent		Feldman, J. L., Kluge, M. D., Allen, P. B.,
                        and Wooten, F., 1993,
        		Phys. Rev. B {\bf 48}, 12589.\\

\noindent		Feldman, J. L., and Kluge, M. D., 1995,
			Phil. Mag. B {\bf 71}, 641.\\

\noindent		Feldman, J. L., Allen, P. B., and Bickham, S. R., 1999,
			Phys. Rev. B {\bf 59}, 3551.\\

\noindent		Feldman, J. L., Bickham, S. R., Engel, G. E., and
			Davidson, B. N., 1998,
			Philos. Mag. B {\bf 77}, 507.\\

\noindent		Gurevich, V. L., 1986,
			{\sl Transport in Phonon Systems},
			Elsevier Science Pub. Co., Amsterdam.\\

\noindent		J\"ackle, J., 1981,
			in {\sl Amorphous Solids: Low Temperature Properties},
			(ed. W. A. Phillips) Springer Verlag, Berlin; p.135.\\

\noindent		Kamitakahara, W. A., Soukoulis, C. M., Shanks, H. R.,
			Buchenau, U., and Grest, G. S., 1987,
			Phys. Rev. B {\bf 36}, 6539.\\

\noindent		Karpov, V. G., Klinger, M. I., and Ignat'ev, F. N.,
			1983, Zh. Eksp. Teor. Fiz. {\bf 84}, 760 [Sov. Phys.
			JETP {\bf 57}, 439.]\\

\noindent		Keating, P. N., 1966,
			Phys. Rev. {\bf 145}, 637.\\

\noindent		Kittel, C., 1948,
			Phys. Rev. {\bf 75}, 972.\\

\noindent		Kubo, R., 1957,
			J. Phys. Soc. Jpn. {\bf 12}, 570.\\

\noindent		Kugler, S., Pusztai, L., Rosta, L., Chieux, P., and
			Bellisent, R., 1993,
			Phys. Rev. B {\bf 48}, 7685.\\

\noindent		Lee, Y. H., Biswas, R., Soukoulis, C. M., Wang, C. Z.,
			Chan, C. T., and Ho, K. M., 1991,
			Phys. Rev. B {\bf 43}, 6573.\\

\noindent		Liu, X., White, B. E. Jr., Pohl, R. O.,
			Iwanizcko, E., Jones, K. M., Mahan, A. H., Nelson,
			B. N., Crandall, R. S., and Veprek, S., 1997,
			Phys. Rev. Letters {\bf 78}, 4419.\\

\noindent		Mott, N. F., and Davis, E. A., 1971,
			{\sl Electronic Processes in Non-Crystalline 
			Materials}, Clarendon Press, Oxford; p.24.\\

\noindent		Nakhmanson, S. M., and Drabold, D. A., 1998,
			Phys. Rev. B {\bf 58}, 15325.\\

\noindent		Oligschleger, C., and Schober, H. R., 1999,
			Phys. Rev. B {\bf 59}, 811.\\

\noindent		Orbach, R., 1996,
			Physica B {\bf 220}, 231.\\

\noindent		Parshin, D. A., 1994,
			Phys. Solid State {\bf 36}, 991.\\

\noindent		Phillips, J. C., 1980,
			Phys. Stat. Solidi (b) {\bf 101}, 473.\\

\noindent               Phillips, W. A., 1972,
 			J. Low Temp. Phys. {\bf 7}, 351.\\

\noindent		Phillips, W. A., 1987,
			Rep. Prog. Phys. {\bf 50}, 1657.\\

\noindent		Pohl, R. O., 1998,
			in {\sl Encylopedia of Applied Physics},
			Wiley-VCH, vol. 23, p.223.\\

\noindent		Pompe, G. and Hegenbarth, E., 1988,
			Phys. Status Solidi B {\bf 147}, 103.\\

\noindent		Schirmacher, W., Diezemann, G., and Ganter, C., 1998,
			Phys. Rev. Letters {\bf 81}, 136. \\

\noindent	   	Schober, H. R., and Laird, B., 1991,
			Phys. Rev. B {\bf 44}, 6746.\\

\noindent		Schober, H. R., and Oligschleger, C., 1996,
			Phys. Rev. B {\bf 53}, 11469 (1996).\\

\noindent		Scholten, A. J., and Dijkhuis, J. I., 1996,
			Phys. Rev. B {\bf 53}, 3837.\\

\noindent		Scholten, A. J., Akimov, A. V., and Dijkhuis, 
			J. I., 1996,
			Phys. Rev. B {\bf 54}, 12151.\\

\noindent		Sette, F., Krisch, M., Masciovecchio, C.,
			Ruocco, G., and Monaco, G., 1998,
			Science {P\bf 280}, 1550.\\

\noindent		Sheng, P., and Zhou, M., 1991,
			Science {\bf 253}, 539.\\

\noindent               Sheng, P., Zhou, M., and Zhang, Z.-Q., 1994,
			Phys. Rev. Letters {\bf 72}, 234.\\

\noindent		Stillinger, F. H., and Weber, T. A., 1985,
			Phys. Rev. B {\bf 31}, 5262.\\

\noindent		Stolen, R. H., 1970,
			Phys. Chem. Glasses {\bf 11}, 83.\\

\noindent		Taraskin, S. N., and Elliott, S. R., 1997,
			Phys. Rev. B {\bf 56}, 8605.\\

\noindent               Thorpe, M. F., 1983,
			J. Non-Cryst. Solids {\bf 57}, 355.\\

\noindent		Trachenko, K., Dove, M. T., Hammonds, K. D., 
			Harris, M. J., and Heine, V., 1998,
			Phys. Rev. Letters {\bf 81}, 3431.\\

\noindent               Visscher, W. M., and Gubernatis, J. E., 1980,
                        in {\sl Dynamical Properties of Solids},
                        (eds.  G. K. Horton and A. A. Maradudin),
                        North Holland, Amsterdam; vol. 4, p.63.\\

\noindent               Weaire, D., and Taylor, P. C., 1980,
                        in {\sl Dynamical Properties of Solids},
                        (eds.  G. K. Horton and A. A. Maradudin),
                        North Holland, Amsterdam; vol. 4, p.1.\\

\noindent		Wooten, F., Winer, K., and Weaire, D., 1985,
			Phys. Rev. Letters {\bf 54}, 1392.\\

\noindent               Zeller, R. C. and Pohl, R. O., 1971,
 			Phys. Rev. B {\bf 4}, 2029. \\

\end{document}